# Polarizable molecular interactions in condensed phase and their equivalent nonpolarizable models

Igor V. Leontyev and Alexei A. Stuchebrukhov*

*Department of Chemistry, University of California Davis, One Shields Avenue, Davis, California 95616*

*E-mails: stuchebr@chem.ucdavis.edu

Earlier, using phenomenological approach, we showed that in some cases polarizable models of condensed phase systems can be reduced to nonpolarizable equivalent models with scaled charges. Examples of such systems include ionic liquids, TIPnP-type models of water, protein force fields, and others, where interactions and dynamics of inherently polarizable species can be accurately described by nonpolarizable models. To describe electrostatic interactions, the effective charges of simple ionic liquids are obtained by scaling the actual charges of ions by a factor of $1/\sqrt{\varepsilon_{el}}$, which is due to electronic polarization screening effect; the scaling factor of neutral species is more complicated. Here, using several theoretical models, we examine how exactly the scaling factors appear in theory, and how, and under what conditions, polarizable Hamiltonians are reduced to nonpolarizable ones. These models allow one to trace the origin of the scaling factors, determine their values, and obtain important insights on the nature of polarizable interactions in condensed matter systems.

# 1. Introduction

At present, most biological simulations utilize nonpolarizable force fields[1, 2]; such are various nonpolarizable models of water (TIP3P[3], SPC[4], more generally TIPnP) and protein force fields: AMBER[5, 6], CHARMM[7], GROMOS[8], OPLS[9, 10], and others. It is recognized however that in some cases these simple force fields give poor results[11] – one example is a low-dielectric protein environment[12, 13], the other is interior of a biological membrane[14]. That is, in some cases nonpolarizable models do produce reasonable results, but in others they do not. The question then can be asked under what conditions an inherently polarizable system can be reduced to a nonpolarizable model with effective and fixed charges, and how to find these effective charges?

A straightforward way to *avoid* such questions is to use polarizable force fields; a number of such force fields are currently being developed [15-26]. Polarizable models, however, still suffer from some unresolved problems, such as consistency and balance in treatment of intra- vs. intermolecular polarizations[27]. The computational efficiency of such models is also an issue. Achieving good sampling of relevant configurations[28] in large biological simulations can become computationally prohibitive if microscopic interactions are computationally expensive[20-25]. Clearly, the simple but computationally efficient nonpolarizable models have advantages over more sophisticated polarizable ones, provided they are accurate enough. Thus, the question is how to derive rigorously an equivalent nonpolarizable model for a polarizable system?

One may also ask if is it possible at all to describe interactions between real molecules, which are inherently polarizable, with a nonpolarizable force field? In general, the answer to this question is negative; e.g. it is not possible to describe both a small cluster of water molecules in vacuum and water molecules in the bulk state with the same set of fixed charges, because in vacuum an isolated water molecule has a dipole 1.85D, whereas in liquid bulk state the dipole is close to 3D[29]. (The exact value of the liquid dipole is still debated; here and below, as previously [29-33], we rely on the results of first principles simulations of liquid state water, Refs. [34, 35]). Thus different thermodynamic states or different environments require different sets of charges to reflect different polarization state in different conditions[36].

However, if simulations involve only configurations that are similar in some macroscopic sense, e.g. as in equilibrium liquid water, there is a well-defined average molecular dipole moment or, more generally, well-defined average charge distribution within the molecule. In this case, a typical set of fixed parameters, such as charges, can be introduced that reflects the averaged values[37]. These averaged charges may even be adjusted to local environment, as in Ref. [36]. The remaining polarization fluctuations around the average can be included as an additional renormalization or scaling of the mean-field parameters [31, 32].

In this paper we consider a generic polarizable model of condensed matter, and examine under what conditions the polarizable Hamiltonian of the system is equivalent to a nonpolarizable one. Starting from an explicitly polarizable model, we will derive an effective mean-field Hamiltonian of an equivalent nonpolarizable model in which the polarization effects are treated implicitly. The formalism is described in the Theory section. The key approximation made is that the electronic polarizability of the system is macroscopically homogeneous, which is usually the case for condensed matter molecular systems, where electronic dielectric constant $\varepsilon_{el}$ is close to 2.0 for most organic materials. The small variations between macroscopic regions such as protein and surrounding water ($\varepsilon_{el}$=1.78) can be included as well. The scaling procedure is similar in spirit to what Warshel introduced in his PDLD/s (scaled PDLD) approach[38]; here we take a closer look at the scaling procedure.

On the one hand, the theory provides theoretical framework for better understanding basic principles of nonpolarizable modeling (of inherently polarizable molecular systems); on the other, it offers insights important for polarizable modeling, suggesting that in some cases a polarizable model (or force field) is exactly equivalent to a nonpolarizable model with scaled charges. The theory provides algorithms for such charge scaling. In such a case, explicit inclusion of polarization dipoles in calculation is not necessary. In the Applications section we discuss some notable examples which show how the scaling principles are applied in polarizable simulations, including models of water.

## 2. Theory

We consider a generic polarizable system, or force field, which consists of point polarizable sites and interaction between them (see e.g. [24, 26]); the interaction includes both electrostatic and non-electrostatic (bonding, van der Waals, etc) terms. Electrostatically, the sites are characterized by fixed point charges and/or point dipoles; the polarization is described by point polarizable dipoles with dipole-dipole interaction between them. We start with a system where all polarizable dipoles interact with each other and all electric sources, and address the issues of neglecting interactions between some neighboring sites, i.e. intra-molecular vs. intra-molecular polarization, later in the section. We also include in consideration the issues of removing, or modifying direct electrostatic interactions between nearest bonded atoms (1-2, 1-3, and 1-4 terms) as such issues appear in standard protein force fields.

Two types of polarizable sites will be considered - polarizable point dipoles and polarizable point charges. Both models are key ingredients in polarizable molecular simulations[26]. For simplify, the higher multipoles[20] are not considered, but the formalism can be extended to such cases as well. The point dipole screening factors are different from those of point charges; the comparison of two models is instructive, as it demonstrates the subtleties of electrostatic screening factors for higher multipoles in comparison with those for point charges. However, in both cases we show that the energy of polarization dipoles (i.e. polarization energy) in the system is proportional to the energy of interaction of the charges or dipoles that induce polarization. This is the key result of the paper that allows one to express the total energy of a polarizable system in terms of the rescaled energy of the underlying system of fixed charges or dipoles.

## 2.1 Polarizable Hamiltonians and equivalent non-polarizable models

Consider a generic polarizable model, which consists of point polarizable sites and interactions between them. The interactions are both electrostatic and non-electrostatic (bonding, van der Waals, etc. [26]). The sites are assigned fixed charges $Q_{0i}$, point dipoles $\vec{\mu}_{0i}$, and polarization dipoles $\vec{\delta}_i$. The interaction Hamiltonian/Energy is

$$W_{tot} = W_x + W_0 + W_{pol} \quad (1)$$

$$W_0 = \frac{1}{2} \sum_{i \neq j} M_{0i}^\sigma K_{ij}^{\sigma\rho} M_{0j}^\rho \quad (2)$$

$$W_{pol} = -\frac{1}{2}\sum_i \vec{E}_{0i}\vec{\delta}_i \qquad (3)$$

The first term $W_x$ includes all non-electrostatic interactions – van der Waals, bonding, torsion, etc.; these terms are not affected by the following transformations, but for completeness we keep them explicitly as they are integral part of any force field. The second term is interaction of fixed charges $Q_{0i}$ and fixed dipoles $\vec{\mu}_{0i}$ on different sites; for each pair of sites there are four terms: charge-charge, dipole-dipole, charge-dipole, and dipole-charge. We write these terms symbolically as interaction of multipoles $M_{0i}^{\sigma} = \{Q_{0i}, \vec{\mu}_{0i}\}$, with corresponding vacuum interaction kernel $K_{ij}^{\sigma\rho}$. For charges the kernel is $K_{ij}^{QQ} = 1/r_{ij}$; for dipole-dipole $K_{ij}^{dd} = (\hat{T}_{ij})_{\alpha\beta} = \dfrac{r_{ij}^2 \delta_{\alpha\beta} - 3(r_{ij})_\alpha (r_{ij})_\beta}{r_{ij}^5}$, etc.

In order to simplify next few transformation steps, we will consider explicitly only the dipole-dipole interaction terms (i.e. one type of interaction instead of four); the results will be easy to generalize when needed. For the dipole-dipole interaction,

$$W_0^{dd} = \frac{1}{2}\sum_{i \neq j}(\vec{\mu}_{0i}\hat{T}_{ij}\vec{\mu}_{0j}) \cdot \eta_{ij} . \qquad (4)$$

To include a possibility that some direct electrostatic interaction terms are excluded or modified, as in 1-2, 1-3, and 1-4 terms in protein fields, we include an additional factor $\eta_{ij}$, which represents modification needed (e.g., $\eta_{ij}=0$, for 1-2, and 1-3 interactions, a constant for 1-4 interactions, and $\eta_{ij}=1$ otherwise). Such terms are also present in the general case of Eq. (2).

Finally, the polarization energy term $W_{pol}$ is written in the form that assumes that the polarization dipoles $\vec{\delta}_i$ are at equilibrium with the local electric field for a given nuclear configuration of the system; the local electric field is due to both fixed charges/dipoles, and other polarization dipoles. The electric field $\vec{E}_{0i}$ is due to fixed charges/dipoles. For a dipoles only system, we have

$$\vec{E}_{0i} = -\sum_j \hat{T}_{ij} \vec{\mu}_{0j}. \qquad (5)$$

$$\vec{\delta}_i = \alpha(\vec{E}_{0i} - \sum_{j \neq i} \hat{T}_{ij} \vec{\delta}_j) \qquad (6)$$

The polarization distribution is found by minimizing total polarization energy for a given nuclear configuration (for discussion, see e.g. Supplementary Information file of Ref. [36] ). Electronic polarization is fast due to high frequency of electronic motion, and therefore it always follows the slow nuclear configuration changes. As nuclei move, the electronic polarization changes adiabatically with nuclei. So, there is an effective "Born-Oppenheimer" potential energy for the nuclei. The problem is how to approximate the effective potential $W_{tot}(r_1, r_2, ... r_N)$; to this end, one needs to evaluate the polarization energy term in Eq. (3). We will do it explicitly for the dipole system, and generalize for other types of interactions later.

Before we move on with formal derivation, one comment regarding fixed parameters $M_{0i}^\sigma = \{Q_{0i}, \vec{\mu}_{0i}\}$ is in order. Although formally the fixed parameters can refer to gas phase state of molecular fragments, and distribution of polarization dipoles $\vec{\delta}_i$ to whatever condensed state we wish, assuming the same force field for both gas and condensed state[20, 26] (a delicate balance between electrostatic and non electrostatic terms is involved), it is also possible to assume a more restricted force field, in which the fixed parameters reflect the average distribution of charges/dipoles in a given condensed state, while polarization describes the remaining fluctuations around the average. (Result of changing fixed parameters, from gas phase to a more relevant condensed state, involves additional energy "re-polarization" terms of Berendsen type of SPC/E model [39], the strain energy due to change of the molecular geometry and other quantum corrections [40], and ubiquitous electronic solvation, see refs [31, 32, 41] for discussion). How to select these average fixed parameters is actually a non-trivial question [19, 24, 42] which goes beyond the scope of this paper; some discussion of re-polarization of water upon transfer from a gas to condensed phase is given in our recent papers picture [31, 36], an additional discussion will be given in the end of this section.

The selection of the fixed parameters also depends on the type of interaction of polarization dipoles with the fixed electric sources in the molecule; one possibility is that polarization dipoles $\vec{\delta}_i$ interact with all sites including own molecular fragment (except of course their own site (i)), another is that interaction with charges/dipoles of a given molecular fragment is set to zero (e.g. for computational reasons). We notice that in the latter case, the gas phase dipole moment of the fragment is defined only by its fixed parameters, while in the former case (interaction with all sites), an additional "self polarization" of the fragments occurs due to interaction with molecular own charges; the dipole moment in the latter case is a combination of fixed parameters, and induced polarization dipoles within the molecular fragment. Clearly, there are many possibilities, and the rational choice of fixed parameters is a subtle non-trivial matter [19, 24, 26, 42]; some of these issues will be discussed in the end of this section.

Here, we will assume first that all polarization dipoles interact with each other, and with all fixed charges, except for those of their own sites. The other possible schemes of interaction will be considered later in the paper.

As seen from Eq. (6), the polarization $\vec{\delta}$ linearly depends on the "external" field $\vec{E}_{0j}$ of fixed charges/dipoles:

$$\vec{\delta}_i = \sum_j \hat{G}_{ij} \vec{E}_{0j} = -\sum_{j,k}(\hat{G}_{ij}\hat{T}_{jk})\vec{\mu}_{0k} = \sum_k \hat{D}_{ik}\vec{\mu}_{0k}, \tag{7}$$

where $\hat{G}$ and $\hat{D}$ are the corresponding Green's functions. We now express the local external field in Eq. (3) in the form of Eq. (5) and arrive at the following expression for polarization energy:

$$W_{pol} = \frac{1}{2}\sum_{ijk}\vec{\mu}_{0i}(\hat{T}_{ik}\hat{D}_{kj})\vec{\mu}_{0j} \tag{8}$$

The polarization energy can now be partitioned into "diagonal" and "non-diagonal" terms:

$$W_{pol} = \frac{1}{2}\sum_i \vec{\mu}_{0i}(\hat{T}\hat{D})_{ii}\vec{\mu}_{0i} + \frac{1}{2}\sum_{i \neq j}\vec{\mu}_{0i}(\hat{T}\hat{D})_{ij}\vec{\mu}_{0j} = W_{pol}^{sol} + W_{pol}^{int} \tag{9}$$

The first term represents the sum of solvation energies of individual sites in the electronic polarizable medium that surrounds each site:

$$W_{pol}^{sol} = \frac{1}{2}\sum_i \vec{\mu}_{0i}(\hat{T}\hat{D})_{ii}\vec{\mu}_{0i} \tag{10}$$

In the condensed state, this *electronic* solvation energy (by its meaning it is negative) is independent of the configuration, e.g. does not depend on the orientation of other dipoles $\vec{\mu}_{0i}$ of the system; it depends only on the properties of individual molecules such as polarizability and average density of polarizable sites in the medium surrounding it, provided that the average polarizability of the medium does not change much as nuclear moves, which is the case for a condensed state. The detailed analysis of the electronic solvation energy of water is given in our recent paper [29].

The second, non-diagonal term in the polarization energy describes the polarization contribution to the effective pair-wise interaction between the sites:

$$W_{pol}^{int} = \frac{1}{2}\sum_{i \neq j} \vec{\mu}_{0i}(\hat{T}\hat{D})_{ij}\vec{\mu}_{0j}. \tag{11}$$

The above expression shows that polarization of the surrounding medium gives rise to an effective interaction of fixed dipoles. This interaction has a typical form of "super-exchange" via an intermediate – polarization dipoles: one dipole induces polarization, which in turn creates electric field that interacts with the second dipole.

It is convenient to add to pair-wise effective polarization interaction of the dipoles in (11) the direct interaction of the dipoles, so that together they would represent the *total* interaction energy of the dipoles in a polarizable medium. The total interaction between dipoles takes the form:

$$W_{tot}^{int} = \frac{1}{2}\sum_{j \neq i}\left(\vec{\mu}_{0i}\hat{T}_{ij}\vec{\mu}_{0j} + \sum_k \vec{\mu}_{0i}(\hat{T}_{ik}\hat{D}_{kj})\vec{\mu}_{0j}\right) = \frac{1}{2}\sum_{j \neq i}W_{tot}^{ij}$$

$$W_{tot}^{ij} = \sum_k \vec{\mu}_{0i}\hat{T}_{ik}(\hat{1}+\hat{D})_{kj}\vec{\mu}_{0j} = \vec{\mu}_{0i}\tilde{\hat{T}}_{ij}\vec{\mu}_{0j} \tag{12}$$

where, $W_{tot}^{ij}$ is the resulting interaction between dipoles in polarizable environment, and

$$\tilde{\hat{T}}_{ij} = \sum_k \hat{T}_{ik}(\hat{1}+\hat{D})_{kj} \tag{13}$$

The functional form of effective interaction is the same as in vacuum, but the interaction kernel $\tilde{\hat{T}}$ is modified compared with the vacuum $\hat{T}$. As for the solvation energy terms, in the condensed state the effective interaction depends mainly on the properties of the electronic polarizable medium represented by the polarizable sites. What is important is that the long-range nature of the electrostatic interactions results in the effective averaging of polarization interactions over large scale, so that only averaged properties of the polarization medium are important. This give rise to the notion of electronic continuum, which is uniform across the condensed matter system, and more or less universal, as the dielectric constant of most bio-organic materials is close to 2.0. Our next goal is to show that in this case the difference between $\tilde{\hat{T}}$ and $\hat{T}$ is a constant, which depends only on the macroscopic dielectric constant of the polarizable medium $\varepsilon_{el}$; therefore total pair-wise interactions are the same as in vacuum, but scaled by a constant,

$$W_{tot}^{ij} = D^{-1}(\varepsilon_{el})W_0^{ij}. \tag{14}$$

The scaling factor depends on the multipole type involved, as we show below.

Summarizing, the results for dipole-type of interaction can be presented as follows:

$$W_{ES}^{dd} = W_0 + W_{pol} = W_0' + W_{tot}^{int} + W_{pol}^{sol} \tag{15}$$

$$W_0' = \frac{1}{2}\sum_{i \ne j}(\vec{\mu}_{0i}\hat{T}_{ij}\vec{\mu}_{0j})(\eta_{ij} - 1) \tag{16}$$

$$W_{tot}^{int} = \frac{1}{2}\sum_{i \ne j}(\vec{\mu}_{0i}\tilde{\hat{T}}_{ij}\vec{\mu}_{0j}). \tag{17}$$

The non-electrostatic terms of $W_x$ remain the same. The electrostatic energy now includes the following terms: 1) modified direct interaction – now it includes the terms that were omitted (due to $(\eta_{ij} - 1)$ factors); 2) the interaction of *all* fixed dipoles in polarizable medium; and 3) solvation energy of fixed dipoles in the polarizable electronic medium.

Under the assumption of macroscopic uniformity of the polarization medium in the condensed state, the last solvation term $W_{pol}^{sol}$ does not depend on the nuclear coordinates, and therefore does not affect dynamics of the nuclei.

The key assumption that we make is that in the condensed state the macroscopic properties of the electronic polarization represented by the polarization dipoles $\vec{\delta}_i$ can be described by a uniform polarizable continuum. The long-range nature of the electrostatic interactions gives rise to an effective averaging of polarization interactions over large scale, so that mostly the large-scale macroscopic properties of the polarization medium are important. The averaging due to nuclear motions also contributes to smearing out the microscopic details of the polarization filed, which also helps to justify the model of a polarizable continuum.

The polarizable properties of the electronic continuum are characterized by the macroscopic electronic dielectric constant of the condensed state material, $\varepsilon_{el}$, which depends only on the average product of the density of polarizable sites $\rho$ and their polarizability $\alpha$ according to Clausius-Mossotti (CM) equation:

$$\varepsilon_{el} = 1 + \frac{4\pi(\rho\alpha)}{1 - \frac{4\pi}{3}(\rho\alpha)}. \tag{18}$$

In the condensed state, when the continuum polarizable medium is formed out of the point polarization dipoles of the Hamiltonian, the resulting electronic dielectric constant of the medium, e.g. calculated from the above equation, may not necessarily be the same as real electronic dielectric of 2.0 or so (for organic materials $\varepsilon_{el}$ are in the range 1.7-2.2 [43]), if the polarization properties of the molecular fragments or sites of the force field were chosen in some formal way to represent the system. (Ideally, of course, the model should reproduce correctly the electronic dielectric constant.) In any case, it is the effective dielectric constant of the polarization continuum formed that will determine the scaling the properties of interactions of a given force field.

Our next step is to express the new terms – the effective interaction and electronic solvation in terms of the continuum model. This will be done in the next subsection.

The expressions in Eq. (15)-(17) represent a principle result, which shows that the system of interacting *polarizable* dipoles $\vec{\mu}_i$ in a condensed phase can be described by an equivalent model of *nonpolarizable* fixed dipoles $\vec{\mu}_{0i}$ with modified interactions. The condensed medium allows for an approximation in which individual polarization dipoles are replaced with a uniform polarizable continuum.

Obviously, the results for dipole-dipole interactions can be generalized to other types of interactions: charge-charge, and charge-dipole. Indeed all the interactions are of the same pair-wise nature, as shown in Eq. (2); the same manipulations can be performed as for the dipoles, and one does arrive at the system of fixed multipoles $M_{0i}^{\rho}$ with modified interactions $\tilde{K}_{ij}^{\rho\sigma}$, which in condensed phase is different from the vacuum interaction by a scaling constant that depends on the electronic dielectric property of the medium. In addition, there are electronic solvation energy terms for a given multipole that do not affect the dynamics but *are* important for solvation energy calculations (see discussion later in the paper).

In the following, we consider different terms of the effective Hamiltonian using simple continuum dielectric models.

## 2. 2. Solvation energies in electronic continuum

Solvation energy terms do not affect dynamics, but are important for solvation energy calculations using polarizable or equivalent scaled models. The solvation energy discussed here refers to individual sites, not to a molecule or molecular fragment as a whole. Solvation energy of a given site charge/dipole is the energy of all polarization dipoles, including site's *own* dipole $\vec{\delta}_i$ in the field of the fixed electric source of the site. For point charges, and point dipoles the problem is easily solved with simple spherical dielectric models considered below; however, when a molecular fragment is treated as a separate unit, several sites need to be considered, including polarization induced interaction between the sites discussed in the next subsection. The problem of total solvation energy of the whole molecule/fragment is more complicated and is not considered in detail here.

It is worth mentioning that a related method for electronic solvation is the usual dielectric continuum solvation model [44, 45], which involves a molecular cavity of arbitrary shape, with point charges/dipoles inside. Here, in addition to a well known problem of defining the molecular boundary, one faces the uncertainty as to how polarization dipoles inside the cavity (self-polarization) should be treated. The models considered below do not address the molecular boundary issue, but provide some insights for self-polarization problem.

For a point charge or a point dipole, the solvation energy can be calculated directly using microscopic expressions such as Eq. (10). In the continuum approximation, these models give the same results as phenomenological Kirkwood-Onsager dielectric model of a spherical cavity in dielectric continuum with a charge or a polarizable dipole in the middle of the sphere.

For a point dipole, as for a charge, the solvation energy depends on the radius of the cavity:

$$W_{di}^{sol} = -\frac{1}{2}\vec{\mu}_{0i}\vec{E}_{RF}(i)$$

$$\vec{E}_{RF}(i) = -\frac{\tilde{\vec{\mu}}_{0i}}{R^3}\frac{2(\varepsilon-1)}{2\varepsilon+1} \qquad (19)$$

$$\tilde{\vec{\mu}}_{0i} = \vec{\mu}_{0i} + \vec{\delta}_i = \frac{\vec{\mu}_{0i}}{1 - \frac{\alpha}{R^3}\frac{2(\varepsilon-1)}{2\varepsilon+1}}$$

There are two possible ways to choose the cavity radius R; both are related to Clausius-Mossotti (CM) equation and to the density of polarizable sites $\rho$:

1. $1/\rho = (2R)^3$, $\quad \dfrac{\alpha}{R^3} = \dfrac{6}{\pi}\dfrac{(\varepsilon-1)}{\varepsilon+2}$ (20)

2. $1/\rho = (4\pi/3)R^3$, $\quad \dfrac{\alpha}{R^3} = \dfrac{(\varepsilon-1)}{\varepsilon+2}$ (21)

In the past, there was an extensive discussion in theory of dielectrics of both models. We previously used the first model, as it better fits the dielectric properties of water. The second choice corresponds to the so-called Lorentz virtual cavity model. This model is consistent with the notion of Lorentz acting field $\vec{E}_{act} = \vec{E}_\varepsilon + 4\pi/3\vec{P}$ that polarizes individual molecules, which is different from the total field in dielectric $\vec{E}_\varepsilon$.

For Lorentz model,

$$\tilde{\vec{\mu}}_{0i} = \vec{\mu}_{0i}\frac{(\varepsilon+2)(2\varepsilon+1)}{9\varepsilon}, \qquad (22)$$

$$\vec{\delta}_i = \alpha\vec{E}_{RF} = \vec{\mu}_{0i}\frac{2(\varepsilon-1)^2}{9\varepsilon} \qquad (23)$$

the solvation energy of a point dipole is

$$W_{di}^{sol} = -\frac{\vec{\mu}_{0i}^2}{2\alpha} \frac{2(\varepsilon-1)^2}{9\varepsilon}. \tag{24}$$

Notice that for a fixed non-polarizable dipole $d_{0i}$, the solvation energy is

$$W_{d0i}^{sol} = -\frac{\vec{\mu}_{0i}^2}{2R^3} \frac{2(\varepsilon-1)}{2\varepsilon+1} = -\frac{\vec{\mu}_{0i}^2}{2\alpha} \frac{2(\varepsilon-1)^2}{(2\varepsilon+1)(\varepsilon+2)}. \tag{25}$$

This energy does not include contributions of the polarization dipole $\delta_i$ on the considered site; the neglect of self-polarization is equivalent to considering a *real* cavity model that often used for calculation of solvation energy.

Comparison of Eq. (24) and (25) shows the difference of two models. The difference is due to a factor in Eq. (22) that relates the self polarized dipole $\tilde{\mu}_{0i}$ to the fixed dipole $\mu_{0i}$ that induces polarization of the surrounding medium, which in turn produces the reaction field that polarizes the dipole $\delta_i$ on the site. For electronic dielectric constant $\varepsilon = 2.0$, the difference in energy is about 10%. Considering that the solvation energy themselves are large in absolute values ($\sim eV$) the difference can be very significant in absolute value.

For a point charge, solvation energy of Lorentz model is

$$W_Q^{sol} = -\frac{Q_{0i}^2}{2R}(1-\frac{1}{\varepsilon}) = -\frac{Q_{0i}^2}{2a}\left(\frac{\varepsilon-1}{\varepsilon}\right)\left(\frac{\varepsilon-1}{\varepsilon+2}\right)^{1/3}, \tag{26}$$

where "polarization radius" is defined as $a^3 = \alpha$. In this case, there is no difference between the real cavity and Lorentz virtual cavity with self-polarization, provided the cavity is spherical and the charge is in the center of the cavity. If this is not the case, the dipole and higher multipoles contributions change the picture and results naturally will depend on self-polarization of the solvated charge.

## 2. 3. Effective interactions and scaling factors

Consider now the interactions between the individual sites. There are tree types: charge-charge, charge-dipole, and dipole-dipole. Again, assuming a continuum polarizable medium formed in the condensed phase, characterized by dielectric constant of Eq.(18), for charge-charge, the interaction scales as

$$W_{int}^{ij} = \frac{1}{\varepsilon} \frac{Q_{0i} Q_{0j}}{r_{ij}} \tag{27}$$

This is total interaction energy of two charges – direct, plus polarization induced part. If direct part was not included in the Hamiltonian initially (e.g. 1-2, 1-3 etc terms), this energy is subtracted in the modified Hamiltonian $W_0\,'$. The effect of polarization can now be described by scaling all charges,

$$Q_{0i} \rightarrow Q_{0i}^{eff} = \frac{Q_{0i}}{\sqrt{\varepsilon}}. \tag{28}$$

For dipoles, the scaling laws are more subtle (details discussed in the Appendix). For the basic model, where all sites have polarization dipoles which interact with all charges on all sites (except its own), the total (direct plus polarization induced) interaction between two fixed dipoles $\mu_{0i}$ and $\mu_{0j}$ has a form:

$$W_{int}^{ij} = \frac{1}{\varepsilon}\left(\frac{\varepsilon+2}{3}\right)^2 \vec{\mu}_{0i} \hat{T}_{ij} \vec{\mu}_{0j}, \tag{29}$$

which means that this interaction can be implicitly included by scaling all dipoles as

$$\mu_{0i} \rightarrow \mu_{0i}^{eff} = \mu_{0i} \frac{\varepsilon+2}{3\sqrt{\varepsilon}}. \tag{30}$$

The scaling factors described above are referred to our basic model, in which individual sites are described by polarizable charges and dipole, and polarization dipoles interact with all electric sources of other sites, except those of its own site. Next we consider a few possible variations of the general scheme.

1. If the model operates with the *total* dipoles, i.e. the fixed dipole plus self polarization dipole, $\tilde{\mu}_{0i} = \mu_{0i} + \delta_i$, which in the Lorentz model is given by (see Eq. (22))

$$\tilde{\mu}_{0i} = \mu_{0i} \frac{(\varepsilon+2)(2\varepsilon+1)}{9\varepsilon}, \tag{31}$$

the total interaction takes the form

$$W_{tot}^{ij} = \varepsilon \left(\frac{3}{2\varepsilon+1}\right)^2 \tilde{\mu}_{0i} \hat{T}_{ij} \tilde{\mu}_{0j}. \tag{32}$$

We see now that the scaling factor for re-polarized total dipoles is different from that in Eq. (29). This factor $D^{-1}(\varepsilon) = \varepsilon \left(\frac{3}{2\varepsilon+1}\right)^2$ is the same as for the model of non-polarizable dipoles (i.e. when a polarization dipole $\delta_i$ is missing on a given site). This is equivalent to having a point dipole in a *real* cavity [31]. In this case we have an effective dielectric with "holes" (because of missing some $\delta_i$), which is quite unnatural, if a continuum dielectric model is considered.

2. Consider a neutral (zero net charge) molecular fragment $(a)$. Suppose there is no interaction of polarization dipoles with the charges of the fragment in the Hamiltonian (but there is interaction with charges of other fragments/molecules, and there is interaction of all polarization dipoles between themselves). Physically it means that the fragment is already self-polarized; the total dipole of the fragment (in gas phase) reflects this self-polarization. Call this molecular fragment dipole $\mu_0(a)$. Consider now a molecular fragment as an individual site. The total dipole moment of the fragment is now equivalent to a fixed dipole $\mu_{0i}$ of a site that does not affect its own polarization $\delta_i$, but can be polarized by other sources, exactly as in our basic model considered above.

The dipolar field outside the fragment can be thought of as a filed of the fixed charges of the fragment and that of self-polarized dipoles that are fixed in the fragment. If a model of a point dipole in a dielectric sphere is applied, the fixed charges dipole $\mu_0(a)'$ and the total dipole $\mu_0(a)$ are related as follows:

$$\mu_0(a) = \mu_0(a)' \frac{3}{\varepsilon + 2} \tag{33}$$

Since the total dipole scales with the factor $\frac{\varepsilon + 2}{3\sqrt{\varepsilon}}$, the scaling factor for fixed charges dipole $\mu_0(a)'$ is:

$$\mu_0^{eff}(a) = \mu_0(a)' \frac{3}{\varepsilon + 2} \frac{\varepsilon + 2}{3\sqrt{\varepsilon}} = \frac{\mu_0(a)'}{\sqrt{\varepsilon}} \tag{34}$$

We see that scaling of the total dipole according to our basic model, is equivalent to scaling fixed charges of the molecular fragment by a ubiquitous factor $1/\sqrt{\varepsilon}$; the individual charges that corresponds to $\mu_0(a)'$ will scale with the same factor $1/\sqrt{\varepsilon}$. Hence the overall procedure of scaling of charges of molecular fragments is as follows:

$$Q_{0i}(a) \to \frac{Q_{0i}}{\sqrt{\varepsilon}} \frac{\varepsilon + 2}{3} \tag{35}$$

This scaling factor is different from a simple $1/\sqrt{\varepsilon}$ as in Eq. (28). The difference is due to specific details of how interactions of the polarization dipoles in the system are chosen.

The relations considered above demonstrate the essence of the approach: in condensed state, all polarization effects are reduced to some sort of scaling of the original fixed charges and dipoles of the model. The charge scaling has been discussed earlier on the basis of much simpler and pure intuitive picture [30, 31, 36]. The present theory provides additional insights and rigor into the question of how the screening factors $D(\varepsilon)$ should be selected.

In the following, we consider some applications of the effective non-polarizable Hamiltonians and discuss their origin and relation to underlying polarizable models.

## 2.4. Molecular Dynamics in Electronic Continuum

*Effective Hamiltonians for MD Simulations.* Summarizing the theory above, we have arrived at the following picture. In the condensed phase, a polarizable force field is (approximately) equivalent to a non-polarizable model of the effective (scaled) fixed charges and dipoles, which reflect implicitly the presence of the electronic polarizable continuum. The model of the charges moving in uniform polarizable continuum is referred to as MDEC – Molecular Dynamics in Electronic Continuum [31-33].

The dynamics of nuclear coordinates in MDEC is expected to be the same as in the equivalent polarizable model; however, the implicit treatment of electronic polarizability requires special care and additional modifications in cases when calculating quantities directly involve electronic polarization. Two such cases are solvation energy calculation, and dielectric constant calculation.

*Free Energy Simulations.* Solvation energy of the entire molecule or a molecular fragment is directly related to solvation energy terms of individual polarization sites in the effective Hamiltonian(15). Overall they contribute to total electrostatic part of the solvation energy related to solvation in electronic continuum $\Delta G_{el}^{*}$. The other part of electrostatic energy is related nuclear configurations of the system, which is explicitly treated in molecular dynamics simulations, $\Delta G_{MD}$. Assuming that MD simulations, and evaluation of energy, is performed by using non-polarizable force field with effective charges, it follows directly from the theory of this paper that the total solvation energy is the sum

$$\Delta G_{sol} = \Delta G_{MD} + \Delta G_{el}, \qquad (36)$$

i.e. the electronic solvation energy needs to added explicitly to solvation energy obtained from MD simulations with non-polarizable force field.

A consistent non-polarizable force field would involve rescaling of all charges in the system by the factor $1/\sqrt{\varepsilon_{el}} \approx 0.7$, including those of ions and ionized groups. However, this is not what is usually done in all known popular force fields; therefore *direct* application of the above equation with such force fields is not possible.

The apparent reason why nonpolarizable force fields such as AMBER[5, 6], CHARMM[7], GROMOS[8], OPLS[9, 10] avoided scaling of the ion charges is that the reduction of charges by itself results in completely wrong hydration free energies[7]. The reason is that the charging free energy of ions from MD simulations by itself does not account for the electronic polarization energy, which typically amounts to more than 50% of the total. This missing electronic energy is usually balanced by using the un-scaled vacuum charges of the ions (and adjusting van der Waals radii), but this balance is not exact, except in high dielectric[32]).

---

[*] Usually, the treatment of solvation energy in terms of continuum models requires consideration of the molecular cavity [46] C.J.F. Bottcher, Theory of Electric Polarization, 2nd ed., Elsevier Sci. Pub. Co., Amsterdam, 1973, [47] V.M. Agranovich, M.D. Galanin, Electronic Excitation Energy Transfer in Condensed Matter, North-Holland, Amsterdam, 1982, [48] M.F. Iozzi, B. Mennucci, J. Tomasi, R. Cammi, Excitation energy transfer (EET) between molecules in condensed matter: A novel application of the polarizable continuum model (PCM), J. Chem. Phys., 120 (2004) 7029-7040.; it should be noticed that molecular cavities *do not* appear in the interaction terms that are related to dynamics, nor do they appear in the solvation energy terms of individual sites.

The problem can be partially resolved by an additional scaling factor to the MD part of free energy, assuming that the main contribution is coming form unscreened charges [30, 31, 36, 49]. Assuming scaling factor $1/\varepsilon_{el}$, the corrected MD part of free energy would be $\Delta G_{MD}/\varepsilon_{el}$ in Eq. (36); one should recognize, however, that the neutral molecular fragments, which are presumably already scaled, in this case become "doubly scaled".

*Dielectric Constant Simulation.* The screened nature of the model charges (and dipole moments) should also be accounted for in simulations of dielectric property. The static dielectric constant is related to the mean square fluctuation of the total dipole $\langle M^2 \rangle$ of the dielectric sample $V$ as[50]

$$\varepsilon_0 = \varepsilon_{el} + \frac{4\pi}{3Vk_BT}\langle M^2 \rangle = \varepsilon_{el}\left(1 + \frac{4\pi}{3Vk_BT}\langle M_{MD}^2 \rangle\right) = \varepsilon_{el} \cdot \varepsilon_{MD} \qquad . \qquad (37)$$

That is the total (static) dielectric constant of the medium $\varepsilon_0$ is a product of that obtained in nonpolarizable MD simulation $\varepsilon_{MD}$ and pure electronic $\varepsilon_{el}$. Here we assumed that all charges were scaled uniformly with a factor $1/\sqrt{\varepsilon_{el}}$. Hence, the MD dipole moment of the sample is related to the actual dipole moment as $\langle M^2 \rangle = \varepsilon_{el}\langle M_{MD}^2 \rangle$, which results in the above relation. The scaling factors between the effective MD dipole moment and the actual dipole moment other than $\sqrt{\varepsilon_{el}}$ are also possible, as discussed in Sections 2.2 and 2.3; however, for the analysis of dielectric data for alkanes discussed below (see Fig. 2), the difference was insignificant, therefore we do not elaborate on these variations further.

***Relation to existing Force Fields.*** With the obtained insight it is interesting to reconsider existing nonpolarizable models and to examine whether they are consistent with the idea of uniform dielectric screening of charges. Parameterization strategy of different force fields is different. It even differs within a force field for different molecules. For example, TIP3P[3] and SPC[4] solvent models were derived completely empirically adjusting their parameters to reproduce in simulations experimental water properties. Empirical parameters of these models obviously should reflect the screened nature of electrostatic interactions. Hence, their empirical charges should be considered as scaled effective charges.

The parameters for other molecules were derived in different procedures. After examining several force fields with very diverse parameterization strategies, we concluded that overall, charges of neutral molecules in common force fields, such as AMBER[5, 6], CHARMM[7], GROMOS[8], OPLS[9, 10], etc, approximately do reflect dielectric screening and can be considered as effective scaled charges.

In contrast, the charges of ionized molecules carry their original un-scaled values ±1 or ±2, as in vacuum, completely disregarding the electronic screening effect inherent to the condensed phase media. Microscopic interaction between such bare charges is obviously overestimated by the factor $\varepsilon_{el} \approx 2$. But this is only half of the problem. The other half is that interactions of ions with solvent are also overestimated. Indeed, if charges of solvent are scaled, while charges of ions are not, then resulting coulomb's interaction is overestimated by the missing factor $\sqrt{\varepsilon_{el}}$ which is about 1.4. To reflect the screened nature of electrostatic interactions charges of ionized moieties need to be corrected by the missing factor $1/\sqrt{\varepsilon_{el}} \approx 0.7$, see additional discussion in refs [31, 32, 41].

## 3. Applications of MDEC model

*3.1. Electronic screening in low dielectric medium.* A remarkable example of how simple charge scaling can replace a fully polarizable simulation is shown in Fig. 1 (initially reported in [32]). The PMF for an ion pair $A^+$ and $A^-$ in benzene was calculated by several methods. VdW parameters of both model ions $A$ correspond to Cl$^-$ atom. In one calculation a fully polarizable CHARMM Drude model [51] was used for benzene (squares in Fig. 1). In another, standard nonpolarizable CHARMM (circles), and finally CHARMM with scaled charges – i.e. MDEC (triangles) was tested. In addition, the least square fitting of the simulation points by the Coulomb function $-1/\varepsilon r$ (with the Ewald correction, see Ref [32]) was done to determine the best empirical dielectric constant for both polarizable and non-polarizable simulations. For polarizable simulations (solid line), the effective dielectric constant was determined to be $\varepsilon_0$=1.88; for nonpolarizable simulations (dashed line), the dielectric constant $\varepsilon_{MD}$=1.16. As seen in Fig. 1, the *nonpolarizable* CHARMM simulations completely fail to reproduce results of polarizable simulations, by up to almost 30kcal/mol in interaction energy; however, the *nonpolarizable* CHARMM simulations with *scaled* charges by a factor $1/\sqrt{(\varepsilon_0/\varepsilon_{MD})}$

(triangles), in accordance with MDEC model, Eq.(37), reproduce results of fully polarizable simulations with remarkable accuracy.

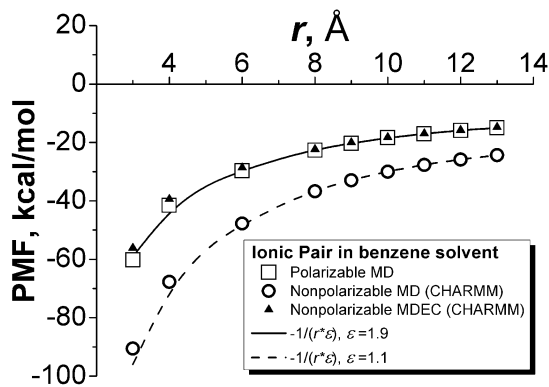

**Figure 1.** PMF of ionic pair in benzene solvent[32], see text for details.

The computed points are accurately approximated by the Coulomb law $1/r$ scaled by a dielectric constant $\varepsilon$, which supports our idea of equivalency of polarizable simulations and charge scaling. The data of polarizable model correspond to the value of dielectric constant 1.9, which exactly represents the electronic polarizability of benzene model in simulations.

*3.2. Calculations of dielectric constants.* The importance of charge scaling (or rather its consequence) is demonstrated next in direct simulated dielectric constants of low-dielectric materials with different methods. Dielectric properties of non-polar alkane series simulated with polarizable CHARMM, non-polarizable CHARMM, and MDEC (non-polarizable CHARM with correction due to charge scaling Eq.(37)) models are shown in Fig. 2. Here MDEC results are obtained from the previously reported data[43]. (The details can be found in ref [33]).

As seen all the values of dielectric constants obtained in standard nonpolarizable CHARMM simulations[43] are about 1.0 (as in vacuum) while experimental values are all about 2. It just means that the standard nonpolarizable model does not capture electronic polarization at all. The problem, however, is efficiently resolved by doing either fully polarizable (and computationally more demanding) simulations as in ref [43] or using effective simple nonpolarizable simulations according to MDEC model.

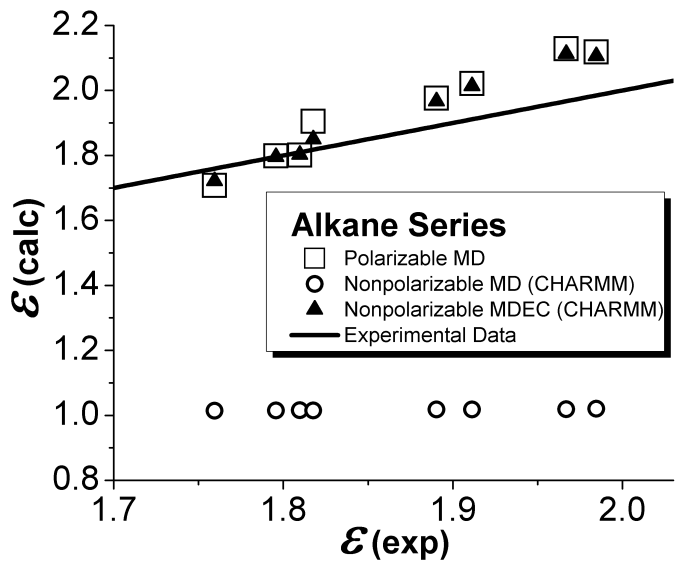

**Figure 2** Dielectric constants of alkane series, see text for details.

It can be argued, of course, that those problems in the low-dielectric solvents are a result of inaccurate CHARMM parameterization that was not calibrated to reproduce solvation and dielectric properties. The CHARMM parameters for alkanes, however, reproduce structural and kinetic properties[43], e.g. rdf and diffusion coefficient, which indicates that microscopic interactions and molecular dynamics are modeled correctly. The over-polarization of the solvent electrostatic parameters for correcting the solvation and dielectric properties would disturb microscopic interactions and result in problems with dynamical properties of the material as it was observed with parameterization of ethers[52]. Moreover, as discussed bellow the over-polarization strategy fails in the case of highly concentrated ionic solutions and ionic liquids. In contrast, with MDEC effective description of the electronic polarization, the problem of compromising between correct solvation and dynamical properties does not arise.

In the case of polar materials such as alcohol series (not shown) the standard nonpolarizable CHARMM model reasonably well reproduces experimental data; yet, the polarizable CHARMM model and nonpolarizable MDEC calculations with Eq. (37) provide much better agreement.

*3.3. Charge scaling of ions.* The charge scaling has received strong support in applications to ions [53-56], in particular in ionic liquids. As reported in ref [53], the use of scaled charges substantially improved the description of dynamical properties of ionic liquids, such as electric conductivity while previously the accurate modeling was reachable only in polarizable simulations[57-60]. The other group[54] investigated very fine effect of interfacial adsorption of halide ions at the oil-water interface. Previously it was believed[61] that the selective effect can be reproduced only in polarizable simulations but in refs [54, 62, 63] the effect was predicted in computationally more efficient nonpolarizable simulations with scaled ionic charges. Another test was to reproduce the neutron scattering structure data for highly concentrated ionic solutions[55]. In standard MD simulations, with unscaled ion charges, the authors observed unphysical clustering of ions due to the exaggerated electrostatic interactions. The correction of ion charges by factor $1/\sqrt{\varepsilon_{el}}$ resolved the problem[55]. In a comparative study[56] of polarizable and non-polarizable models, the nonpolarizable

MDEC simulations with scaled charges demonstrated capability of reproducing ion pairing structure properties. Overall, there is growing evidence in the literature [53-56] that the lack of electronic polarizability in nonpolarizable simulations can be efficiently compensated by simple scaling of ionic charges.

*3.4. Solvation free energy.* The idea of charge scaling of ions, although justified physically, sometimes is rejected in practice on the basis of apparent difficulty[7] of reproducing the ion hydration free energy (main target property for parameterization of ions), as the reduced charge significantly decreases the solvation energy of MD simulation in comparison with experiment. The issue is naturally resolved within the MDEC framework, in which the reduced free energy from MD simulations is compensated by the electronic free energy term, as given by Eq. (36). As shown in Fig. 3, hydration free energies for polyatomic ions are remarkably well reproduced in MDEC simulations[49].

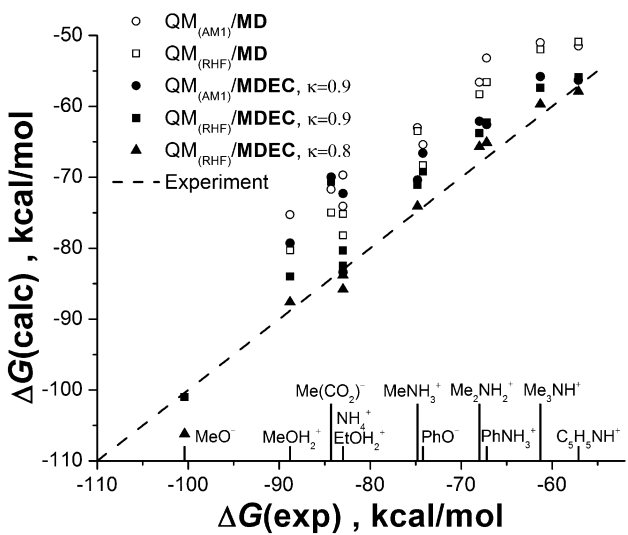

**Figure 3** Charging free energy of polyatomic ions in aqua solution[31, 49].

In the Figure, opened symbols correspond to standard MD simulations with GROMACS force field[8] (G43A); filled symbols stand for MDEC simulations, which used scaled charges of ions and Eq.(36). The experimental values (dashed line) were obtained as a difference of total solvation energies for a given ion[64] and its hydrocarbon counterpart[65]. Charges and geometry of ions were found in the SCRF/ESP quantum procedure[66]. Circles correspond to AM1 level of theory, while squares and triangles to RHF/6-31G**. Van der Waals radii scaled by the factor $\kappa$=0.9 (filled circles and squares) and $\kappa$=0.8 (filled triangles) were used to build the dielectric cavity in MDEC computation of $\Delta G^{el}$ term. The results demonstrate that the new technique reconciles the long standing issue of inconsistency of the correct scaled ionic interactions and correct hydration free energies.

## 4. Conclusions

In this paper we examined the formalism for the implicit treatment of polarizability in computer simulations. Starting from the potential energy expression for explicitly polarizable model we have derived an effective mean-field Hamiltonian of nonpolarizable model. In other words, the analysis represents a formal proof that under certain conditions, inherently polarizable molecular systems can be described by equivalent *nonpolarizable* fixed-charge models. This is the main result of the present paper.

The theory states that if simulations involve only similar configurations so that there is a well-defined average molecular dipole moment or, more generally, a well-defined average charge distribution within the molecule, then a typical set of fixed parameters, such as charges, can be introduced that represent an averaged charge distribution[37]. We showed that the remaining fluctuations of charges around the average can be included as an additional renormalization or uniform scaling of the mean-field parameters[31, 32]. The renormalization of the original mean-field charges is equivalent to a familiar screening by the electronic polarizable continuum[19,20] with dielectric constant $\varepsilon_{el}$ (also known as optic dielectric constant $\varepsilon_\infty$), which is a measurable characteristic for a given condensed state. For example, for liquid water the electronic part $\varepsilon_{el}$ of the dielectric constant is 1.78, and for most organic materials this value is close to 2.0. The resulting fixed-charge model is equivalent to the developed earlier MDEC model [31-33] (Molecular Dynamics in Electronic Continuum). Minor variations of the electronic dielectric constant between macroscopic regions such as protein and outside solvent, can be easily addressed within the general framework of continuum dielectrics.

The model involves only effective scaled charges, and does not include electronic polarizability explicitly. The electronic polarizable continuum, however, is a part of the model, and has to be included explicitly in certain cases, especially when a molecule is transferred from a condensed phase to gas phase, such as in calculation of solvation free energy[30] or vaporization energy[29], and also dielectric properties[33] of the material. As in our previous works, the theory states that charges of ionized groups of the protein, as well as charges of ions, in nonpolarizable simulations should be scaled in respect to the actual mean-field values; i.e. reduced by a factor $1/\sqrt{\varepsilon_{el}}$ (about 0.7), to reflect the electronic screening of the condensed phase medium. In the solvation free energy simulations the electronic part of the free energy (estimated by the continuum model) should be added explicitly to the nuclear reorganization part, which is obtained in nonpolarizable MD simulations.

The present theory also sheds new light on the effect of dielectric screening of polar molecules which have been debated in the literature for a long time [30, 31, 47, 48, 67-70]. It was shown that screening of charges for neutral molecules can vary from $1/\sqrt{\varepsilon_{el}}$ (about 0.7) to a much weaker factor ~0.9. The exact value of the factor depends on the model of *intramolecular* polarizability which can not be defined unambiguously even in most sophisticated polarizable models [20-25].

In a few examples, we compared the traditional nonpolarizable MD simulations with MDEC approach, and demonstrated how the new principles improve the accuracy of the computationally-efficient nonpolarizable approach. The successful examples of MDEC applications are encouraging, but it is of course recognized that simple scaling is not a replacement of a well-build real polarizable force field. It is hoped, however, that the obviously useful charge scaling idea will find its proper place in the future force field models.

## Acknowledgements

This work has been supported in part by the NSF grant PHY 0646273, and NIH GM054052.

# Appendix

**Scaling factors in Lorentz virtual cavity model**

Consider a probe point dipole $\mu_0$ in dielectric $\varepsilon$. The field of the dipole scales as [46]

$$E_\varepsilon \propto \mu_0 \frac{\varepsilon+2}{3\varepsilon} . \qquad (A1)$$

This is the scaling factor for the field of a point dipole (i.e. probe nonpolarizable dipole $\mu_0$) in polarizable medium. It is obtained as follows. Suppose there is a point charge $q$ in the medium that creates electric field $E_\varepsilon$; the field in the medium scales as $q/\varepsilon$. The acting field on the dipole is then modified by Lorentz virtual cavity as [46]

$$E_{act} = E_\varepsilon \cdot \frac{\varepsilon+2}{3} \qquad (A2)$$

Therefore energy of interaction of the charge and the probe dipole is

$$W_{int} \propto q \frac{\varepsilon+2}{3\varepsilon} \mu_0 . \qquad (A3)$$

Due to symmetry of electrostatic interactions, the same interaction can be obtained as the energy of the probe charge $q$ in the field of the dipole $\mu_0$. Hence the field created by the probe dipole scales as shown in (A1).

Now if the interaction of probe dipole $\mu_{0i}$ with another probe dipole $\mu_{0j}$ is considered, the acting field is

$$E_{act} = E_\varepsilon \cdot \frac{\varepsilon+2}{3} \propto \mu_{0j} \cdot \frac{1}{\varepsilon}\left(\frac{\varepsilon+2}{3}\right)^2 . \qquad (A4)$$

Hence interaction of two probe dipoles $\mu_{0i}$ and $\mu_{0j}$ is

$$W_{int}^{ij} \propto \mu_{0i} \frac{1}{\varepsilon}\left(\frac{\varepsilon+2}{3}\right)^2 \mu_{0j} . \qquad (A5)$$

In the condensed phase medium, however, the nonpolarizable probe dipoles $\mu_{0i}$ and $\mu_{0j}$ are in some sense unobservable. It is sometimes more convenient to consider re-polarized dipoles $\tilde{\mu}_{0i}$ and $\tilde{\mu}_{0j}$, which correspond to *total* dipole moments of molecules in the medium, i.e. dipoles that include part of the medium polarization induced by the probe dipoles themselves. The relation between probe $\mu_0$ and re-polarized $\tilde{\mu}_0$ dipole can be obtained in the Kirkwood-Onsager (KO) model[71, 72]:

$$\tilde{\mu}_0 = \frac{\mu_0}{1 - \frac{2(\varepsilon-1)}{(2\varepsilon+1)}\frac{\alpha}{R^3}}, \qquad (A6)$$

where $R$ is the radius of molecular cavity in dielectric of $\varepsilon$, and $\alpha$ is the molecular polarizability. The most natural choice for the cavity radius $R$ with respect to polarization density $\rho$ of the CM equation:

$$\frac{4\pi}{3}R^3 = 1/\rho. \qquad (A7)$$

This choice corresponds to the Lorentz virtual cavity model. In this case the relation between the total self-polarized dipole $\tilde{\mu}_0$ and $\mu_0$ is

$$\tilde{\mu}_0 = \mu_0 \frac{(\varepsilon+2)(2\varepsilon+1)}{9\varepsilon}. \qquad (A8)$$

Now we can rewrite the interaction energy (A5) in terms of re-polarized dipoles. This interaction, when scaling (A8) is included, takes the form:

$$W_{\text{int}}^{ij} \propto \tilde{\mu}_{0i}\, \varepsilon \left(\frac{3}{2\varepsilon+1}\right)^2 \tilde{\mu}_{0j}. \qquad (A9)$$

The scaling factor in Eq. (A9) is equivalent to that of the *real* cavity model[31].

Previously, many authors [47, 67, 68, 70] considered scaling that follows from Eq.(A5); it should be recognized that it corresponds to scaling of un-polarized dipoles $\mu_{0i}$ and $\mu_{0j}$, while for total self-polarized dipoles, the scaling factor is different, Eq. (A9).

# Figure captions

**Fig. 1** PMF for an ion pair $A^+$ and $A^-$ in benzene.

**Fig. 2** Dielectric constants for alkane series computed with different models. The squares and circles stand for polarizable and standard nonpolarizable CHARMM simulations, respectively, taken from ref[43], (see also ref [73] for alcohol series). The triangles represent nonpolarizable MDEC calculations, Eq. (37), where $\varepsilon_{MD}$ and $\varepsilon_{el}$ are taken from ref [43]. The solid line stands for the experimental data[74].

**Fig. 3** Charging free energy of polyatomic ions in aqua solution simulated in the ref[49]. Opened symbols correspond to the standard MD simulations, while filled symbols stand for MDEC technique, eq. (37).

**Figure 1**

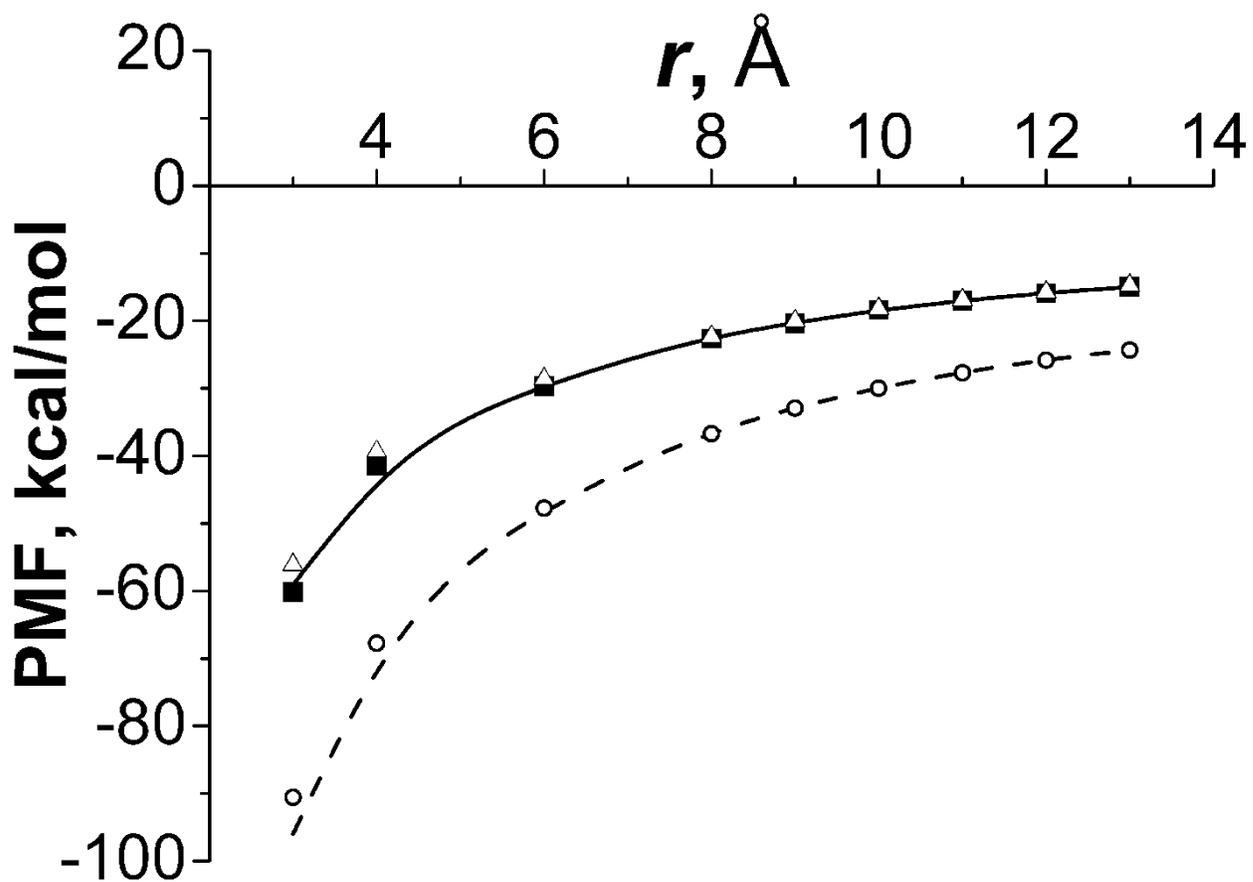

**Figure 2**

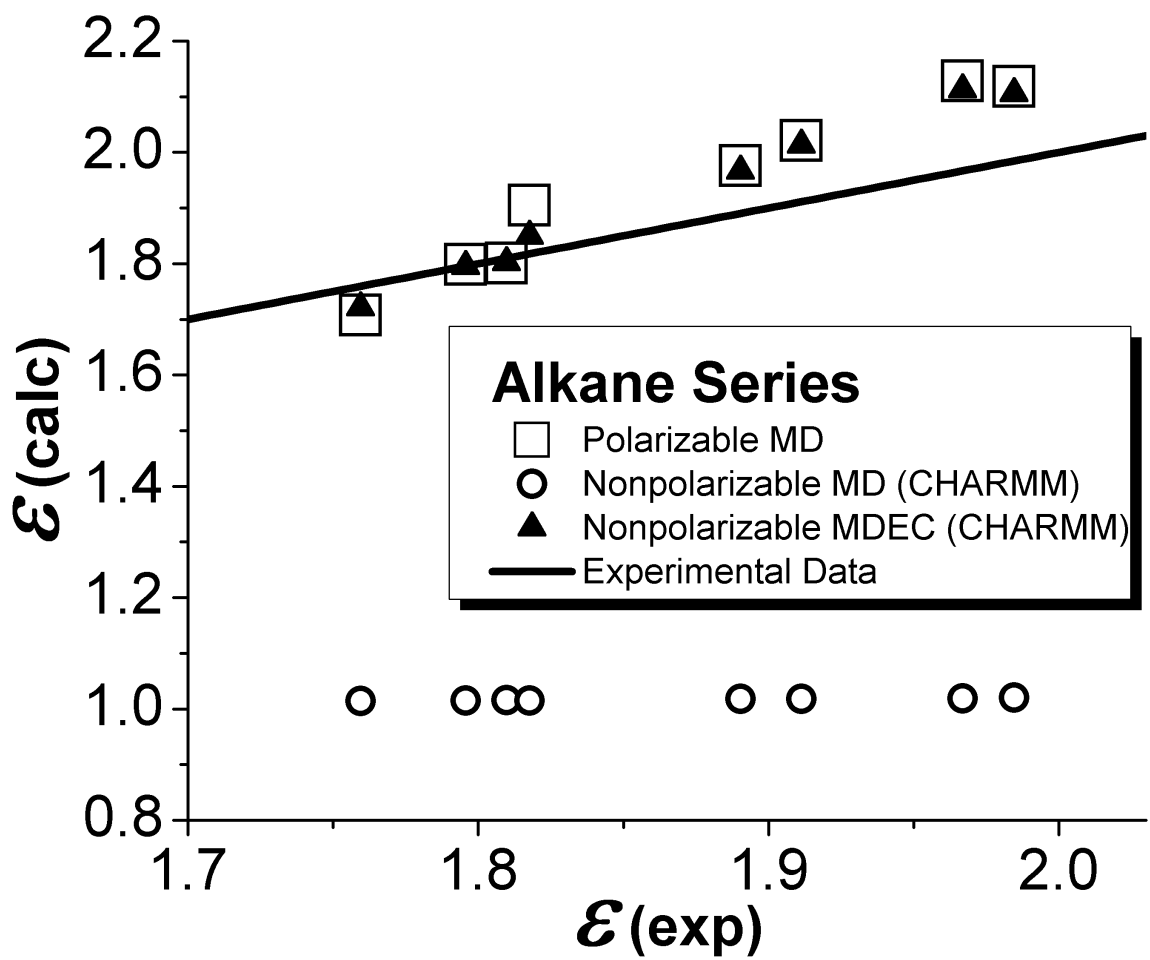

Figure 3

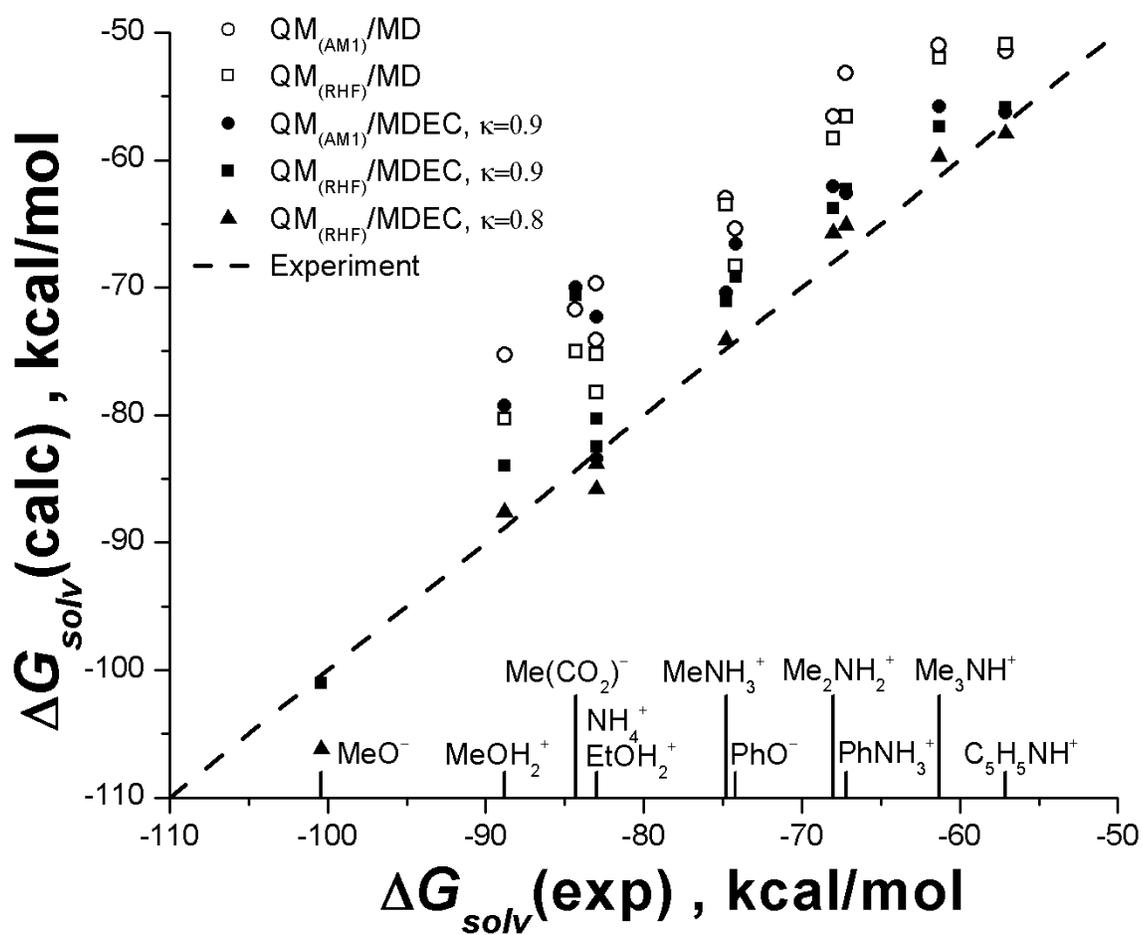